\documentclass[12pt]{article}                  

\usepackage{graphicx,amsmath,amssymb,amsfonts,latexsym,color,dcolumn,bm}
\usepackage{braket}
\usepackage{makeidx}
\usepackage{multirow}
\usepackage[dvipsnames,svgnames,table]{xcolor}
\usepackage{graphicx}
\usepackage{epstopdf}
\usepackage{ulem}
\usepackage{titlesec}

\RequirePackage{geometry}
\RequirePackage{graphicx}
\RequirePackage{color}
\RequirePackage{cite}

\begin{document}
\newcommand{\be}{\begin{equation}}
\newcommand{\ee}{\end{equation}}
\newcommand{\bea}{\begin{eqnarray}}
\newcommand{\eea}{\end{eqnarray}}
\newcommand{\ad}{a^{\dag}}
\newcommand{\la}{\langle}
\newcommand{\ra}{\rangle}
\newcommand{\om}{\omega}
\newcommand{\Ep}{E^{(+)}}
\newcommand{\Em}{E^{(-)}}
\newcommand{\pa}{\partial}

\newcommand\mytitle[1]{ {\LARGE \bf 
  \begin{center} #1 \end{center}}\vskip-0.5cm }
\newcommand\myauthor[1]{ {\vskip0.1cm\noindent{
  \large\begin{center} \bf #1 \end{center}}\vskip-0.4cm}}
\newcommand\myaddress[1]{ {\vskip-0.5cm\noindent
  \begin{center} {\it #1} \end{center} \vskip-0.4cm} }
\newcommand\myemail[1]{ {\vskip-0.5cm\noindent
  \begin{center}{ \rm #1 } \end{center} \vskip-0.5cm} }
\renewcommand\maketitle{\vskip0pt}

%Title of paper
\mytitle{Complementarity in single photon interference - the role of the mode function and vacuum fields}
\myauthor{R. Menzel}
\myaddress{University of Potsdam, Institute of Physics and Astronomy, Karl-Liebknecht Stra\ss e 24-25,
 D-14476 Potsdam}
\myemail{menzel@uni-potsdam.de :corresponding author}
\myauthor{D. Puhlmann}
\myaddress{University of Potsdam, Institute of Physics and Astronomy, Karl-Liebknecht Stra\ss e 24-25,
D-14476 Potsdam}
\myemail{dirk.puhlmann@uni-potsdam.de}
\myauthor{A. Heuer}
\myaddress{University of Potsdam, Institute of Physics and Astronomy, Karl-Liebknecht Stra\ss e 24-25,
D-14476 Potsdam}
\myemail{heuer@uni-potsdam.de}

\maketitle

\textbf{Abstract :}
Single photon first order interferences of spatially separated regions from the cone structure of spontaneous parametric down conversion allow for analyzing the role of the mode function in quantum optics regarding the complementarity principle. In earlier experiments the role of the vacuum fields could be demonstrated as the source of complementarity with respect to the temporal properties \cite{Heu14}. Here the spatial coherence properties of these vacuum fields are demonstrated as the physical reason for complementarity in these single photon quantum optical experiments. These results are directly connected to the mode picture in classical optics.\\


\section{Introduction}

Complementarity is one of the most important principles of quantum physics \cite{Bohr28}. It is directly connected to the “measurement problem” \cite{Glaub}. In quantum optics light is detected as “clicks” of the single photons, which correspond to the transfer of an energy packet of $h v_{photon}$ to the detector, with the spatial, temporal, spectral and polarization filter restrictions of the detector setup. The measured intensity of bright light results from the statistical superposition of these ``clicks''. The possible positions of these ``clicks'' are given by the interference pattern of the measured light fields. For a single photon this interference is given by the overlay of the electric field components in the right order of the field operators belonging to this photon at the detector ${\la E_{photon}^{(-)}E_{photon}^{(+)}\ra}$ \cite{Glaub}. The possibly coherent light fields produce interference fringes resulting in certain visibility $V$. In some cases the sources of the photons can be distinguished resulting in ``which-path'' distinguishability $D$. The complementarity principle in quantum optics answers the question: How coherent are the more or less distinguishable shares of the electric field (or how distinguishable are the paths of the more or less coherent light modes)? The upper limit for the combined measurement of both is  ${D^2+V^2 \leq 1}$ \cite{Jaeger95}. The visibility is based on the coherence of the involved light modes in the different photon paths and is calculated from the maximum and minimum single photon count rates S of the interference fringes by ${V = (S_{max}-S_{min})/(S_{max}+S_{min}) }$. It can be concluded that coherence in single photon measurements decreases if paths become distinguishable \cite{Man91}. Thus the question about the physical details behind this quantum law of complementarity may be asked.\\

In a previous set of experiments we investigated the physical background of complementarity in the temporal domain and showed the role of involved vacuum fields \cite{Heu14}. Using spontaneous parametric down conversion (SPDC) in induced coherence experiments first-order interference visibilities of more than 95 \% were observed if the single photon was emitted potentially from two different SPDC-crystals, indistinguishable. And, as expected, the visibility dropped to zero if the single photon path, which means the source of emmission, could be determined, and vice versa. Because these photons were generated by three-wave-mixing of a coherent pump field with certain vacuum fields a deeper physical analysis was possible. High visibility could be obtained for the single photon interference, only, if the same vacuum field was acting in both crystals and no fixed phase relation occurred between the two photon waves if the involved vacuum fields could be distinguished. In this case the two waves of the single photon“emitted synchronously by the two crystals” even in the same TEM$_{00}$-mode  are not coherent in contrast to the classical expectation. In summary, the random phase relations between distinguishable vacuum fields were identified as physical reason for the complementarity principle in the temporal dimension in this case. \cite{Heu14}. Based on these results the relation of coherence and distinguishability for single photons in the spatial dimension is investigated, here.\\ 

Quite some work was done to investigate the coherence properties of the emitted light from SPDC. For example the coherence area of the emission was investigated with the result that the source is incoherent across the pump spot indicating a thermal emitter \cite{Rib94}. This fulfills expectation because SPDC is a spontaneous process. A similar but more detailed result was worked out experimentally and theoretically \cite{Joob94, Joob96}. In \cite{Joob96} it was shown that especially in radial direction coherence effects from phase matching and other details provide a more complex than thermal behavior of this SPDC light. In all these investigations the correlation between the coherence of the signal and idler photons was analyzed. These correlation and entanglement properties were also investigated in detail in \cite{Just13, Paul13, Jost98}. The spatial emission can be nicely described using the Schmidt mode decomposition as investigated  in \cite{Exter09, Kulik09, Exter12, Leuchs15}. But only very few investigations deal with the coherence of the single photons which is the aim of this work (see e.g. \cite{Rib94, Joob94}).\\

\begin{figure}[h]
\includegraphics[width=12cm]{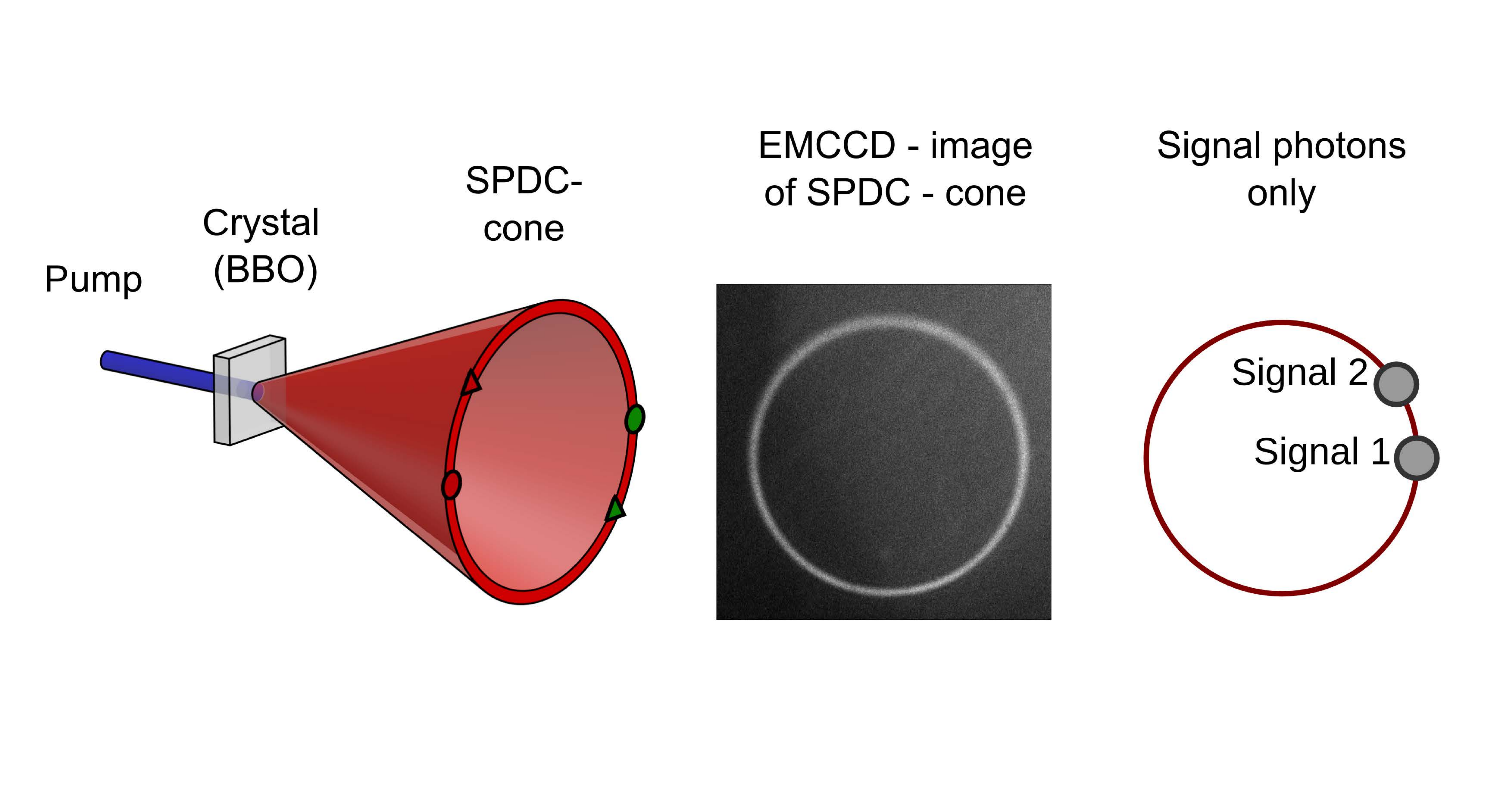}
\caption{Scheme of the emitted light cone in type I SPDC: The entangled signal (red) and idler (green) photons occur at opposite sides of the cone (left side of the figure). In the middle of the figure a photo of the cone emission is shown. It is investigated, how the visibility (coherence) decreases if the interfering TEM$_{00}$-modes signal 1 and signal 2 of a single photon are separated within the light cone e.g. tangentially as shown on right of the figure. }
\label{Fig1}
\end{figure}

\begin{figure}[h]
\includegraphics[width=12cm]{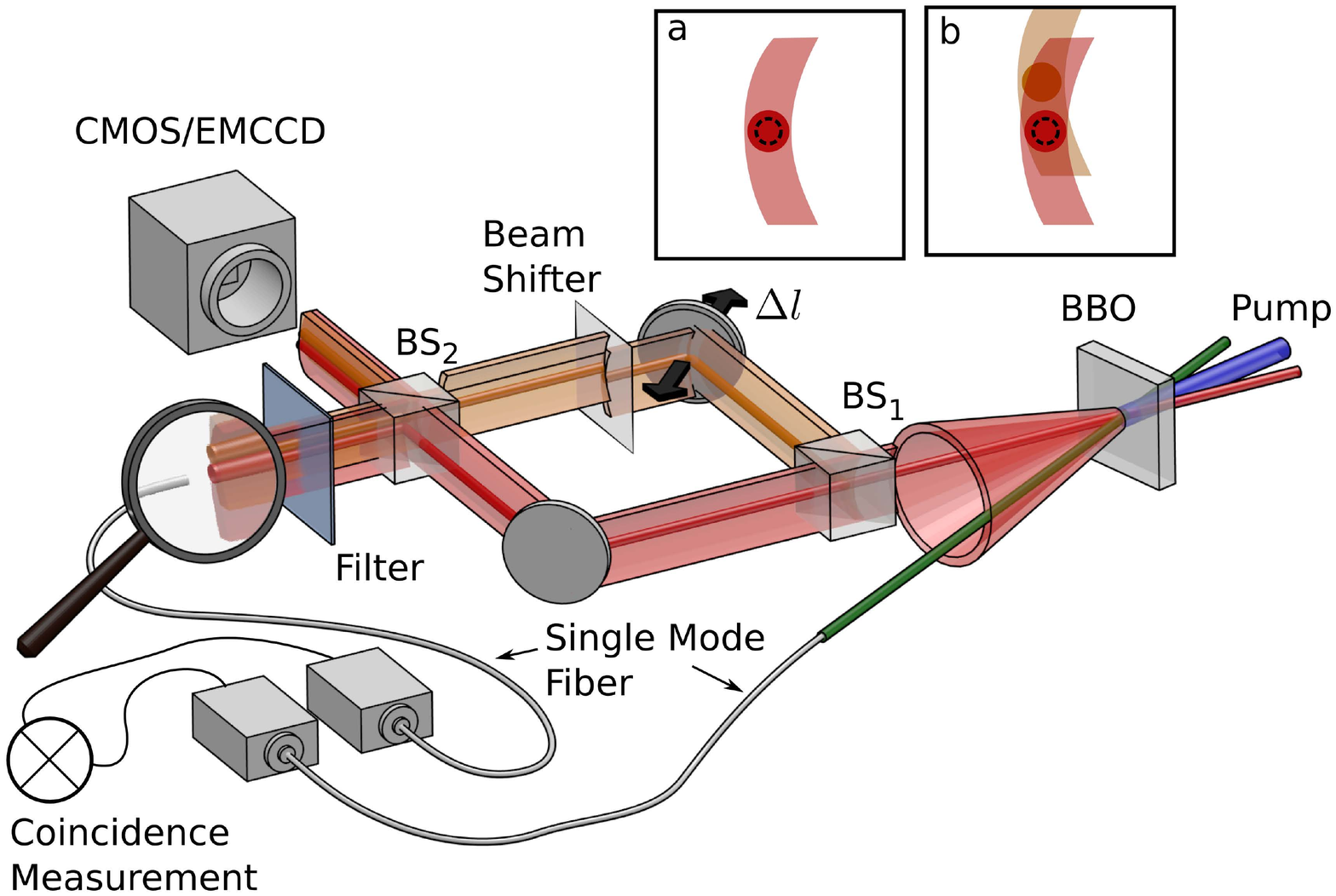}
\caption{Experimental setup: For parametric down conversion a 2 mm long BBO crystal was used. The crystal was cut for collinear type I phase matching around 405 nm. The path delay between the two arms of the Mach-Zehnder-interferometer for observing the fringes and determining the visibility was realized by a motorized high resolution translation stage ($\Delta$l). The beam selection from different positions of the light cone structure as schemed in inset b was realized by a beam tilt and shift arrangement which also allowed for the perfect overlay of the selected two modes at the second beam splitter BS$_2$. The detectors are single mode fiber coupled avalanche photo diodes SPCM-AQRH-13 from Perkin Elmer with lenses in front for determining the TEM$_{00}$ –modes of the idler- and the two signal-fields.}
\label{Fig2}
\end{figure}

Therefore we investigated the transversal coherence of single signal photons generated via spontaneous parametric down conversion (SPDC) by measuring them with a TEM$_{00}$-mode detector in two separated modes as illustrated in Fig. \ref{Fig1} and \ref{Fig2}. In type I SPDC the two entangled signal and idler photons have opposite momentum components with respect to the pump photon and appear diagonally within the emmission cone as illustrated in Fig. \ref{Fig1}. The selection of a single fundamental Gaussian mode (TEM$_{00}$) is always possible for any light source (see e.g. \cite{Men98, Kulik09, Leuchs15}). Using two tilted arms of a Mach-Zehnder interferometer we measure the relative coherence of two tilted and on top of each other realigned Gaussian modes of a single signal photon selected out of the light cone of SPDC which are emitted from the same volume of the crystal (see Fig. \ref{Fig1} and \ref{Fig2}).\\

As demonstrated in the theoretical section the single photons in the selected TEM$_{00}$-detector modes are generated by three-wave-mixing of the coherent pump beam and a this way selected vacuum field. Therefore the coherence properties of these photons are directly dependent on the coherence properties of the involved vacuum fields. As result the properties of the vacuum fields can be identified as the physical background of the complementarity principle in the spatial dimension, in this case. The details will be discussed in the conclusion.\\

\section{Experimental}
The experimental setup is shown in  Fig. \ref{Fig2}. The 2mm long BBO crystal as nonlinear material was pumped by a diode-laser (Blue mode, Toptica) with the emission peak wavelength of 405 nm and a cw power of 30 mW. From the emitted light cone the signal-photons were selected on one side of the cone and fed into the Mach-Zehnder-interferometer at the first beam splitter BS$_{1}$. In one arm of the interferometer a beam shifter was used to tilt and shift one of the interfering TEM$_{00}$-modes (signal 2 in Fig. \ref{Fig1}) radially and tangentially relative to the reference TEM$_{00}$-mode (signal 1 in Fig. \ref{Fig1}) in the other interferometer arm. Both modes were perfectly overlaid in the second beam splitter BS$_{2}$  to match the TEM$_{00}$-mode of the detector. With this TEM$_{00}$-mode detector the single photon interference of the two beams is measured. This detector is constructed with a transversal single-mode fiber and an aspherical lens in front of it. It is a selector for all photons belonging to the TEM$_{00}$-mode of the fiber, only. The beam waist of this mode was aligned to match the TEM$_{00}$-mode of the pump spot in the BBO crystal by size (165 $\mu$m in diameter) and position. Both have a Gaussian beam profile. With a delay $\Delta$l realized with a linear translation stage (Newport) allowing longitudinal delays of 25 mm with an accuracy of 10 nm the interference fringes between signal 1 and signal 2 were measured and the visibility calculated.\\

The resulting interference pattern was photographed with the EMCCD camera (see Fig. \ref{Fig2}) on the other side of the beam splitter cube BS$_{2}$. At this position also the tangential and radial distance between signal 1 and 2 was measured using a not shown HeNe-laser and a CMOS camera instead of the EMCCD.\\

A spectral filter was applied in front of the single mode fiber with a spectral band width of 2.5 nm (FWHM) at a peak wavelength of 808 nm. For decreasing the signal to noise ratio of the visibility measurement the signal photon interference of first-order could also be measured in coincidence with the temporally correlated idler photons on the other side of the cone resulting in visibilities of almost one. But all visibility measurements given here were performed as single photon measurements without any corrections and not in coincidence.\\

While the position of the idler photon detection was kept fixed at the opposite position of signal beam 1, a coincidence measurement between signal and idler photons allowed the determination of the distinguishability ${D = (R_{path 1}-R_{path 2})/(R_{path 1}+R_{path 2}) }$. ${R_{path 1}}$ and ${R_{path 2}}$ are the coincidence rates of the idler photons with photons of signal 1 and signal 2, respectively.\\

\section{Results and Discussion}
\begin{figure}[h]
\includegraphics[width=12cm]{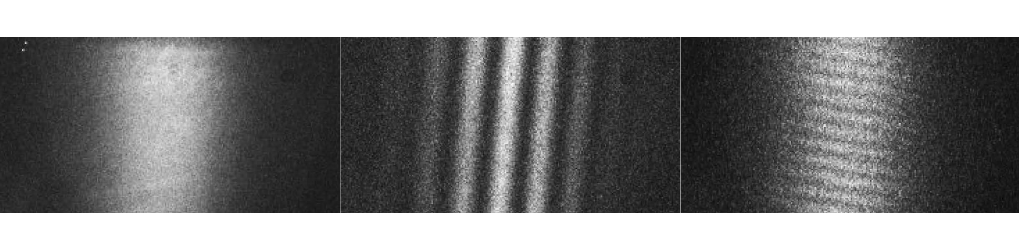}
\caption{Three light patterns photographed with the EMCCD-camera behind the single signal photon Mach-Zehnder interferometer for illustration of the decrease of coherence while tilting one of the two measured TEM$_{00}$ signal modes. Perfect alignment (left), with a slight horizontal angle between the beams (middle). Tilt of one of the beams results in a loss of fringe contrast and visibility (right).}
\label{Fig3}
\end{figure}

Perfect alignment of the Mach-Zehnder interferometer results in an unstructured bright pattern from the overlaid same share of the cone indicating perfect local coherence (see photo in Fig. \ref{Fig3} (left)). Aligning both beams in a very small angle in almost horizontal direction results in fringes from the first-order interference of the single signal photons with high contrast as can be seen in the middle trace of Fig. \ref{Fig3}. Tilting the two interferometer arms tangentially (as in Fig. \ref{Fig3}, right) or radially with respect to each other results in a decrease of the fringe visibility. The fringe distances and directions result from the different angles in the alignments with no effect to the fringe visibility.\\

\begin{figure}[h]
\includegraphics[width=12cm]{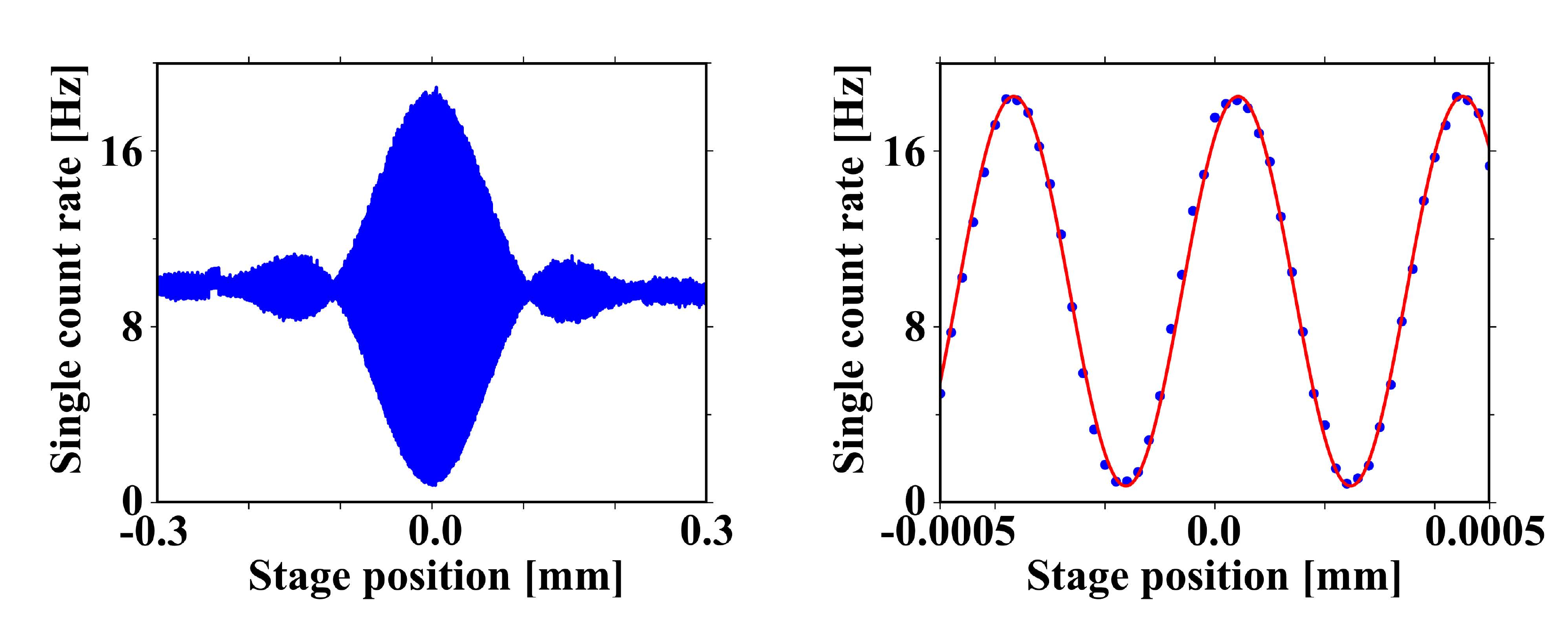}
\caption{Measurements of the Mach-Zehnder interference fringes with the single photon detector for perfect alignment as a function of the position of the delay line with different resolutions.}
\label{Fig4}
\end{figure}

All the following quantitative measurements are done with the best possible alignment of the two modes with respect to the detector mode as in Fig. \ref{Fig3} (left) and the fringe visibility was determined by moving the translation stage for longitudinal delays ($\Delta$l in Fig. \ref{Fig2}). In Fig. \ref{Fig4} the single photon interference pattern of the signal photons as a function of the translation stage position (longitudinal delay) is shown while overlaying the same spot of the light cone (signal 2 is the same mode as signal 1). It was measured with the single photon detector behind the second beam splitter. In the left graph of this figure the result of a low resolution measurement is given. It allows the determination of the longitudinal coherence length l$_c$ of the measured signal photons to 83 $\mu$m (half width at half maximum of the interference signal), which is in good agreement with the bandwidth ${\Delta \lambda}$ of 2.5 nm (FWHM) of the applied spectral interference filter (${l_c = \lambda^2/(\pi \cdot \Delta \lambda)}$).\\

\begin{figure}[h]
\includegraphics[width=5.5cm]{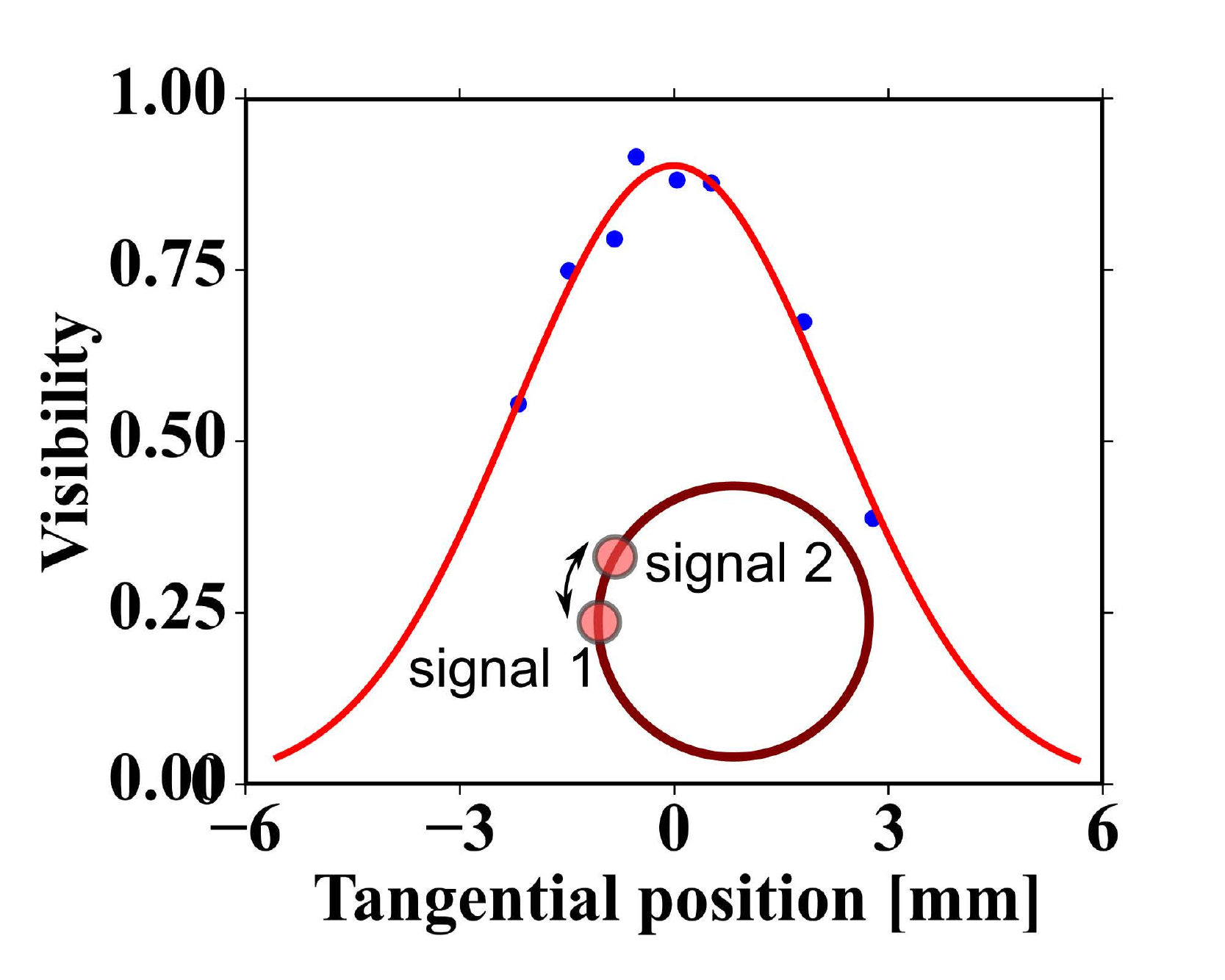}
\includegraphics[width=5.5cm]{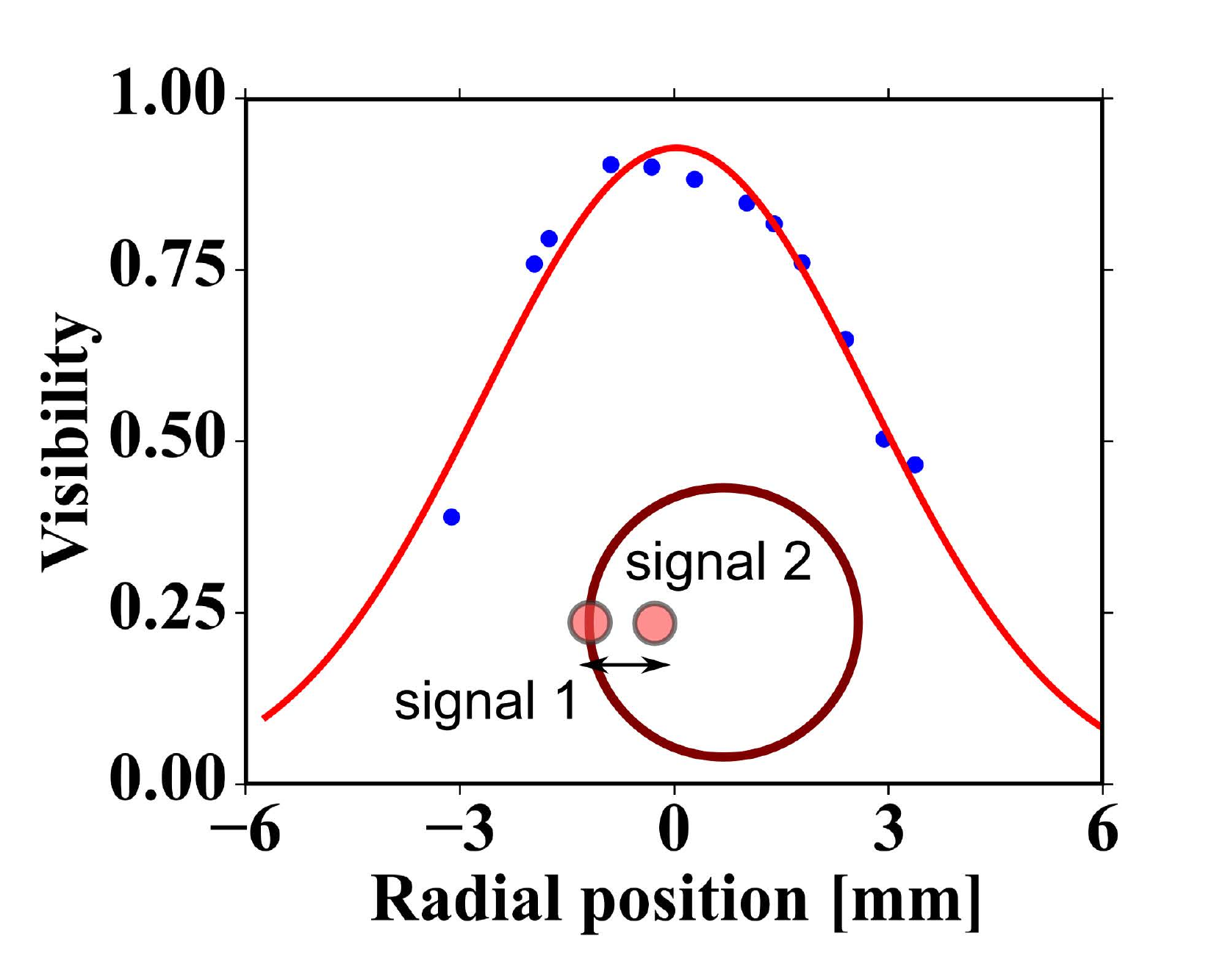}
\caption{Visibility as a function of the relative tilted and thus in the reference plane shifted interfering TEM$_{00}$-modes of the single signal photons in tangential and radial direction. The solid line is the result of the best fit of the experimental data using a Gaussian distribution in agreement with the theoretical model.}
\label{Fig5}
\end{figure}

In the graph of Fig. \ref{Fig4} (right) the result of a high resolution measurement for the same alignment is given. The maximum observed visibility of the not corrected single photon measurements is 90 \% in both cases. This visibility measurement is used as a reference for the following sets of experiments where one of the Mach-Zehnder interferometer arms is tilted transversally in tangential and in radial direction. The visibility for each data point was measured in the same way as in Fig. \ref{Fig4}. The decrease of the visibility is evaluated as a function of the tilts. The results of these measurements are depicted in Fig. \ref{Fig5}.\\

In Fig. \ref{Fig5} (left) the result of the visibility evaluation is given as a function of a tangential tilt of one beam (signal 2 in Fig. \ref{Fig5} and Fig. \ref{Fig1}) with respect to the fixed one (signal 1). No tilt results in maximum visibility of about 90 \% as given in Fig. \ref{Fig5}. But visibility drops rapidly if the mode of signal 2 is tilted more than half the divergence angle of the selected Gaussian beam. It has to be pronounced that both interfering beams are well within the light cone structure and thus they both have almost undiminished intensity. This can be seen in Fig. \ref{Fig3}, right, which is a measurement with less than 40 \% visibility. The tilt of the beam in radial direction results in a similar value as can be seen in Fig. \ref{Fig5} (right). For comparison with the theoretical calculations the $1/\sqrt{e}$-half-widths of the fitted visibility measurements of Fig. \ref{Fig5} are ($2.2\pm0.3$) mm in tangential and ($2.7\pm0.3$) mm in radial direction and the associated maximum visibilities are 0.90 and 0.93.

\begin{figure}[h]
\includegraphics[width=5.5cm]{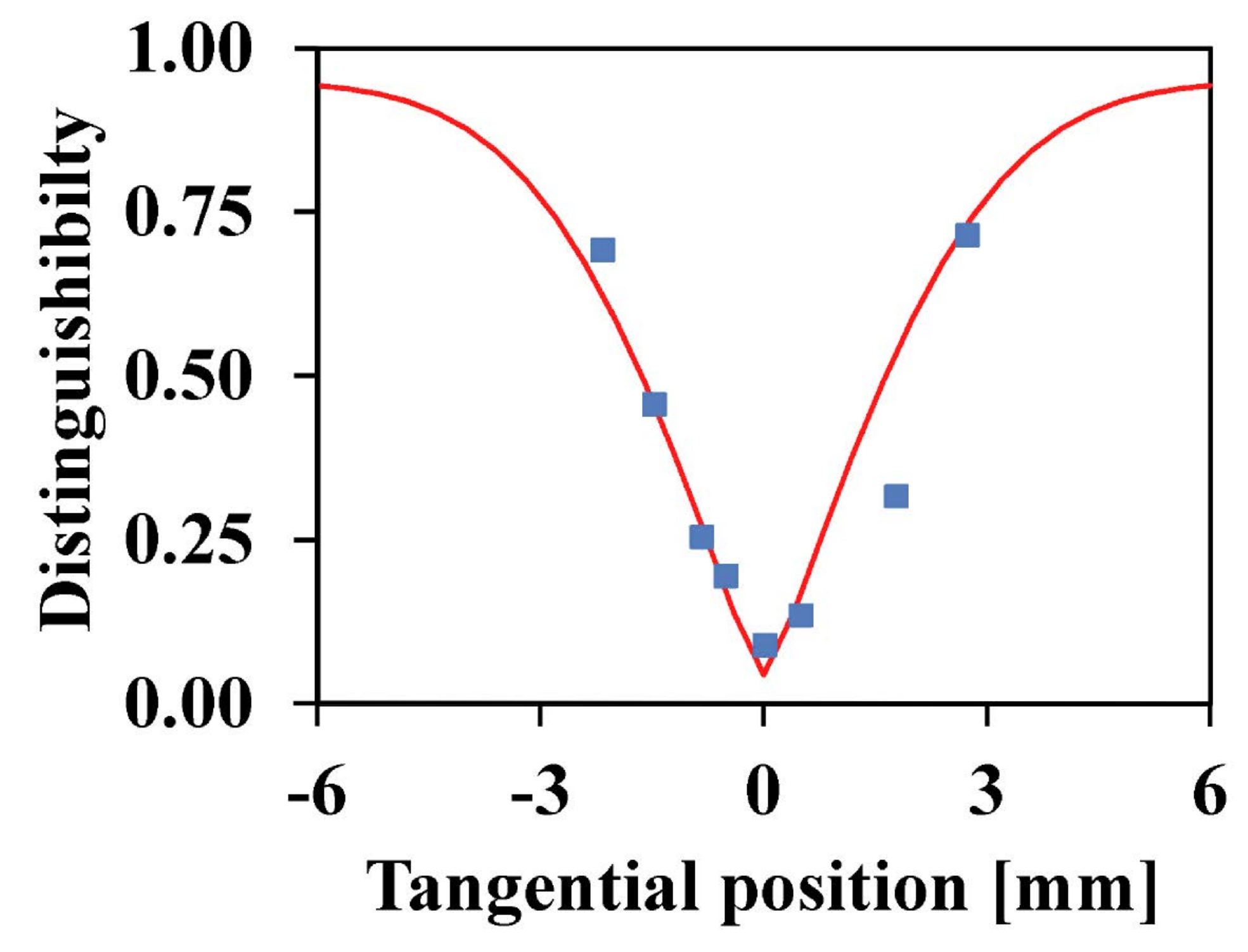}
\caption{Distinguishability of the single signal photons in the two tilted and thus in the reference plane shifted TEM$_{00}$-modes in tangential direction and measured in coincidence with the reference idler photon. The solid line is a fit of the experimental data using the result of Fig. \ref{Fig6} (left) as reference.}
\label{Fig6}
\end{figure}

For the measurement of figure 5 left the distinguishability D of the two paths of the single signal photons in the two TEM$_{00}$-modes of signal 1 and signal 2 was measured in coincidence with the idler photon as reference (see Fig. \ref{Fig1} and Fig. \ref{Fig2}). The result is given in figure \ref{Fig6}. As expected the distinguishability is minimal in case of overlapping modes and increases as the visibility decreases. It will approach 1 if the modes are completely separated. In between the ${D^2+V^2}$ expression has experimental values of about 0.8-0.9. The limitation below 0.9 can be understood as the result of the maximum visibility of 0.9. The two experiments of  Fig. \ref{Fig5} (left) and Fig. \ref{Fig6} demonstrate the complementarity principle for the single signal photons, directly. It has to be pronounced that in all these measurement no background corrections have been applied.

\subsection {Theoretical description}
The theoretical analysis of the experimental data can be based on a simplified model as applied in \cite{Heu14, Mil96} using an effective Hamiltonian with couplings $a_P\ad_S\ad_I$ and $\ad_Pa_Sa_I$, where $a_P,a_S$, and $a_I$ ($\ad_P,\ad_S$, and $\ad_I$) are photon annihilation (creation) operators for the pump, signal, and idler fields. This describes the annihilation (creation) of the pump photon and the simultaneous creation (annihilation) of signal and idler photons in the SPDC 3-wave mixing process with the vacuum field contributions ${a_{S0 1}}$ and ${a_{S0 2}}$.

We assume perfect phase matching and restrict this analysis to spectral single field modes of frequency $\om_S$ and $\om_I$, respectively, which is experimentally realized by the narrow spectral filter. The pump is treated as an un-depleted, classical field. From the Heisenberg equations of motion for the field operators we write the photon annihilation part of the total electric field operator for the two relevant signal field modes selected by the detector as
\bea
E_{S 1,2}^{(+)}(\bold r, t)&=&\big[ a_{S0 1,2} +C \ad_{I0 1,2} \big]U_{S0 1,2}(\bold r) e^{-i (\bold k_{1,2}\bold r+\om_St)}
\label{th1}
\eea
ignoring an irrelevant multiplicative constant. C is a constant that incorporates the crystal properties and the classical pump amplitude. The strong spatial TEM$_{00}$-mode filtering by the detector in this experiment extracts small spatial shares from the emitted field, only. Therefore in addition to the usual description as in \cite{Heu14, Mil96} spatial mode functions for the two measured TEM$_{00}$ signal field modes propagating in the directions $\bold k_{1,2}$ through the interferometer are included as $U_{S 1,2}$ as e.g. described in \cite{Ling08}. These directions are chosen by the position of the detector behind the two interferometer arms (see Fig. \ref{Fig2}) and thus the momentum conservation and phase matching angular restrictions of the down conversion process are considered. The mode functions $U_{S 1,2}$ for the two TEM$_{00}$-detector-modes are given by
\bea
U_{S0 1,2}(x,y,z)&=&\frac{1}{1-i \frac{z \lambda}{w_0^2 \pi}}e^{-i \bigg[ \frac{x^2+y^2}{w_0^2 \big(1-i \frac{z \lambda}{w_0^2 \pi}\big)}\bigg]}.
\label{th1}
\eea
 At the second beam splitter BS$_2$ the two fields are superimposed as shown in Fig. \ref{Fig2}. In the upper arm a phase shift $\phi$ and the beam shifter for the TEM$_{00}$-mode of signal 2 are implemented. Thus behind the second beam splitter BS$_2$ the electric field is given by
\bea
E_S^{(+)}(\bold r, t)&=& E_{S1}^{(+)}(\bold r, t)+E_{S2}^{(+)}(\bold r + \Delta \bold r, t) e^{-i\phi}
\label{th1}
\eea
where $\Delta \bold r$ takes care of the tilted and thus in the reference plane shifted TEM$_{00}$-mode (signal 2 in Fig. \ref{Fig1}). 

Assuming low conversion efficiency, we retain only the lowest-order terms which is single bi-photon generation. The signal photon counting rate in the selected mode of the detector as given in Eq.2 is proportional to 
\bea
R_{detector}&=&\int_{-\infty}^\infty \int_{-\infty}^\infty \la E_S^{(-)}E_S^{(+)}\ra dx dy
\label{th1}
\eea
As result the interference visibility V at the reference plane can be calculated by the normalized cross correlation of the two here in y-direction shifted transversal field distributions of the two modes of signal 1 and signal 2 (see Fig. \ref{Fig1}).  
\bea
V(\Delta y) &=&\frac{ 2 \int_{-\infty}^\infty e^{-\frac{y^2}{w^2}} e^{-\frac{(y+ \Delta y)^2}{w^2}}dy}{ \int_{-\infty}^\infty e^{- 2 \frac{y^2}{w^2}} dy+\int_{-\infty}^\infty e^{-2 \frac{(y+ \Delta y)^2}{w^2}}dy}\\
&=&e^{-\frac{\Delta y^2}{2 w^2}}
\label{th1}
\eea
In this equation the mode selection is represented by the value of $\Delta y$ which contains the propagation directions of the two TEM$_{00}$-modes and thus the geometry restrictions of the involved vacuum fields, too. The beam radius w is the field radius given by the detector mode radius w$_D$ at the crystal, the propagation distance z from the crystal to the reference plane and the wavelength $\lambda$.
\bea
w = w_D^2 \sqrt{1 +\bigg(\frac{z \lambda}{w_D^2 \pi}\bigg)^2}
\label{th1}
\eea
The optical length from the crystal to the reference plane was 661 mm and thus the coherence radius of the observed mode was 2.06 mm at the reference plane. This radius is identical with the $1/\sqrt{e}$-half-width of the theoretical visibility curve. This is in agreement with the fit of the experimental data of this complex measurement. In other words, the observed transversal coherence width of the single photons is as large as the width of the selected detector mode for the single photons, although the total intensity distribution is much larger. For the tangential displacements it was 2.22 mm resulting in a relative error of 7\%. The theoretical result for the radial displacements shows a relative error of about 25\%, which may still support the fundamental idea of this work but in contrast in the radial direction phase matching effects may play a role \cite{Joob96}, which are not considered here.\\

\section{Conclusion}
Although the emitted light cone structure is much wider than the observed mode, the transversal coherence length of the single photons within the cone is just as large as the detected single photon mode. This result seems anti-intuitive because especially in tangential direction, along the ring structure, no symmetry breaking feature, as e.g. from the phase matching conditions, limits the transversal coherence along the ring. The emission of a single photon in each of the observed directions is equally probable.\\

The only limiting factor to explain this experimental result is the selection of the involved modes. The analysis as it was applied in \cite{Heu14} shows that the 3-wave mixing process for generating the single signal photons involves besides the coherent pump light also vacuum fields. By the  observed TEM$_{00}$-detection-modes the relevant vacuum field modes are selected as TEM$_{00}$-mode structures, as well. The tilt of one of the photon modes is associated with a tilt of the related vacuum field mode (see Eq. 3 and 5). This tilted vacuum field has a random phase compared to the non-tilted vacuum field and the visibility drops rapidly whereas the distinguishability increases.\\

In other words with this experiment the influence of the involved vacuum fields regarding spatial coherence is observed. From this experiment it can be stated that a single photon TEM$_{00}$ mode is coherent. This is also observed with bright light in classical optics. But using single photons the distinguishability of the different photon paths could be measured, also. We measured that distinguishable other modes of this photon are not coherent to this. In experiments with higher order modes as e.g. TEM$_{01}$ \cite{Elsn15} this analogy could be observed indirectly, too. Photons from the same coherent mode are not distinguishable and distinguishable photons belong always to different modes. From these experiments and the results of \cite{Heu14} it can be concluded that vacuum as under laying background for all photon generation processes may introduce general randomness in quantum optics in the spatial as well as in the temporal dimension. But by selecting coherent vacuum fields in the photon generation process we see coherent photons allowing interference experiments with visibilities of up to 1. This interplay is the physical background of the complementarity principle in quantum optics and it is the result of the mode selection in the measurement process.\\

\textbf{Declarations}\\

\textbf{Acknowledgment}\\
We gratefully acknowledge P. W. Milonni for very intense and illuminating discussions and the long-lasting collaboration.
We also like to thank M.W. Wilkens always having an open ear for us.\\

\textbf{Competing interests}\\
The authors declare that they have no competing interests.\\

\textbf{Authors’ contributions}\\
RM proposed the original idea, gave advices and wrote the manuscript. DP implemented the work. The progress is a result of common contributions and discussions of RM and AH. All three authors read and approved the final manuscript.

\textbf{Funding}\\
No funding was received

\textbf{Figure captions}\\

\end{document}